\documentstyle[preprint,aps,prc]{revtex}

\begin{document}
\title{Temperature and energy partition in fragmentation}
\author{A. Strachan and C. O. Dorso}
\address{Departamento de F\'\i sica, Facultad de Ciencias Exactas 
y Naturales,\\
Universidad de Buenos Aires, Pabell\'on 1, Ciudad Universitaria, Nu\~nez\\
1428, Buenos Aires, Argentina.}
\maketitle

\begin{abstract}

We study fragmentation of small atomistic clusters via molecular dynamics.
We calculate the time scales related to fragment formation and
emission.
We also show that some degree of thermalization is achieved during the
expansion process, which allows the determination of a local temperature.
In this way we can calculate the break-up temperature as a function
of excitation energy, i.e. the fragmentation caloric curve. 
Fragmentation appears as a rather constant temperature region
of the caloric curve. Furthermore, we show that different definitions
of temperature, related to different degrees of freedom, yield
very similar values.

\end{abstract}


\section{Introduction}

The highly non equilibrium process of fragmentation, i.e. the 
expansion and break-up of an excited system, remains of great 
importance both from the theoretical and experimental points of
view. In the field of nuclear physics heavy ion collisions at 
intermediate energy provide one of the most important examples of 
this phenomena \cite{pocho97}, but the process of fragmentation is also 
important in other branches of physics, for instance: fragment 
formation in ion beam or laser ablation sputtering experiments 
\cite{sputt}, fragmentation of fluids flowing 
through nozzles, etc. In spite of the great efforts done in order 
to characterize the process of fragmentation and to understand the 
mechanisms and instabilities that lead to the break-up, several 
aspects remain obscure. Most noticeable, the possible relation 
of this process with the liquid-gas phase transition or 
critical phenomena are, at present, not clearly understood. 

One of the main present lines of research is focused on the
determination of the temperature of the fragmenting system at break-up 
time which, when analyzed in terms of the excitation energy leads to the 
the nuclear fragmentation caloric curve. 
The meaning of temperature and the interpretation of the caloric curve 
in fragmentation processes face two important problems. On the one
hand the systems are usually very small ($\sim 100$ constituents).
This is not a serious problem, at least for low excitation energies
when the small system is self-confined; it has been recently shown that 
very small systems can undergo phase transitions \cite{labastie90} and 
calorimetric experiments have been done in systems as small as 
$139$-particle Na clusters \cite{calorimetria}.
The second problem is related to time, the expanding system and 
the clusters are transient systems. This is the same problem
that appears when studying isolated liquid clusters; fortunately 
the metastable states usually live long enough to be measured.
This said, let us mention some of the recent efforts to calculate
the fragmentation caloric curve, i.e. temperature as a function of
energy.
Pochodzalla et al. \cite{pocho95} have calculated the ``isotope 
temperature'' as a function of the total energy per nucleon in 
different nuclear reactions. They calculated the ``isotope temperature''
in experiments performed by the ALADIN
collaboration. Isotope temperature is computed from ratios
of clusters yields, assuming that the system is at low density
and that thermal and chemical equilibrium is achieved \cite{albergo}.
In the fragmentation region they find two behaviors:
in the energy range $3$ MeV to $10$ MeV the temperature remains
almost constant with a value of $\sim 5$ MeV. The temperature value
of this plateau coincides
with the limiting excitation energy per nucleon for fusion-evaporation
process. For excitation energies beyond $10$ MeV they find a steady 
increase of the isotope temperature. Comparing this caloric curve
with the one characteristic of liquid-gas phase transition in the
thermodynamic limit, the behavior
for high energies, was related to the presence of the
nuclear ``gas phase'', the region of increasing temperature, for high
energies in the fragmentation caloric curve
will be called ``gas branch''. In numerical simulations, using
a quantum molecular dynamics model for the study of fragmentation 
\cite{bondorf}, a similar behavior is found for the temperature of the 
gas of fragments, although some criticisms can be raised against the later 
calculations regarding the way the collective expansion is taken
into account.
Recently the isotope temperature was calculated by the EOS collaboration
for Au on C central collisions. The plateau is confirmed but the
rise at high energies, i.e. the ``gas branch'', is not found \cite{hauger}.
In very recent experiments \cite{serfling} the ``emission or
excited-state temperature''
was calculated over a wide excitation energy on Au+Au reactions.
Contrary to the results is \cite{pocho95}, the emission temperature
is rather constant throughout the whole energy range, and does not
increase for high energies. For an excellent review on the calculation
of the caloric curve in nuclear reactions and the different definitions
of temperature see \cite{pocho97}. In simulations of fragmentation of
two dimensional classical systems via molecular dynamics, the
fragmentation regime appears, also, as a flat region in the
caloric curve \cite{caloric}. It is thus very important to
understand the fragmentation caloric curve, in particular
to understand its high energy behavior, and check whether
the ``gas phase'' behavior, i.e. the gas branch, is seen or not.

In this paper we study the fragmentation of excited classical Lennard-Jones
(L-J) drops, formed by $N = 147$ particles, simulated via molecular dynamics. 
We chose L-J pair potential because it is a general potential featuring
short-range repulsion and longer-range attraction, which is needed
for clusterization to happen. The equation of state of L-J systems displays 
Van der Waals behavior as is expected for nuclear systems.
Fragmentation in L-J drops has been studied extensively by many
authors \cite{pandha85,pandha86,pandha87,times}, this simple model
has given a lot of information about the process of fragmentation.
Understanding the process of fragmentation in classical LJ drops
does not mean understanding nuclear multifragmentation, but
it is unlikely that we will understand the nuclear process
if we do not understanding it in the simple LJ case.

The main purpose of this work is to show that we can define a ``local
temperature'' of the fragmenting system which achieves some degree of 
local equilibrium after a very short period of time. Also, we developed
a way to 
define precisely the time at which the fragments are formed. 
Thus we can calculate the break-up temperature in fragmentation,
from which we can calculate the caloric curve
of our L-J system in a very wide energy range, encompassing the solid-like and
liquid-like phases (and the associated phase transition) and, more important
to us, evaporation and fragmentation processes.

In order to unveil the mechanisms that lead to the
fragmentation of an expanding system, it is mandatory to know
the times at which fragments are formed and emitted. 
We pay particular attention to the calculation of the 
time-scales related to cluster formation and emission, as we will see
this is important to the physical meaning of the ``local temperature''.

Another characteristic feature of fragmenting systems 
is the development of a collective expansion or radial flux. Identifying
this collective motion correctly, is essential to calculate temperature.
We study the way in which the total energy is partitioned
into potential energy ($V$), collective kinetic energy ($K_{coll}$), 
associated to the radial flux, and internal kinetic energy ($K_{int}$)
as a function of time and for different total energies of the system.
We will then see how two local temperatures can be defined, one
related to the fluctuations of the velocity of the particles 
over the radial, collective motion, and the second related to the
fluctuations of the center of mass velocities of the clusters.
The internal temperature of the clusters is also calculated as
a function of the fragment mass, for different times. The relation
among the different temperatures is analyzed.

This paper is organized as follows: in section II we will describe 
the model that we use to simulate the fragmentation process. In 
section III is devoted to the study of fragment formation and
emission time-scales,
followed by our results on the properties of the asymptotic fragments
in section IV. Section V deals with the energy partition, and in
section VI we show the caloric curve for our Lennard-Jones system 
in a wide energy range encompassing fragmentation and also
the solid-like and liquid-like phases. Finally, in section
VII, conclusions are drawn.

\section{Computer experiments}

As stated in the Introduction we study fragmentation of excited
Lennard-Jones drops. The two body interaction potential is taken as
the truncated Lennard Jones (6-12) potential:

\begin{equation}
V(r) = \left\{  \begin{array}{ll}
4\epsilon \left[ \left( \frac \sigma r\right) ^{12}-\left( \frac \sigma
r\right) ^6-\left( \frac \sigma {r_c}\right) ^{12}+\left( \frac \sigma
{r_c}\right) ^6\right] & r < r_c \\
0                      & r \ge r_c 
		\end{array}
	\right.
\end{equation}

We took the cut-off radius as $r_c=3\sigma$. 
Energy and distance are measured in units of the potential well 
($\epsilon $) and the distance at which the potential 
changes sign ($\sigma$), respectively. The unit of time used is: 
$t_0=\sqrt{\sigma ^2 m/48\epsilon }$. We used the well known
Verlet algorithm to integrate the classical equations of motion 
\cite{frenkel} taking $t_{int} = 0.001 t_0$ as the integration
time step. This led to energy being conserved approximately
one part per million.

We performed explosions of $N=147$ particles, three dimensional drops.
The initial configurations are constructed by cutting a spherical drop 
from a thermalized, periodic, Lennard-Jones system with $N = 512$
particles in each periodic cell. The degree of excitation can be
easily controlled in this way by varying the density and temperature
of the periodic system. The initial
state of our drops is macroscopically characterized by their energy and
density (taken as that of the periodic system). We studied a broad energy
range which encompasses very different behaviors regarding the fragmentation
pattern, going from $E = -2.4 \epsilon$ to $E = 2.2 \epsilon$. The density
was taken as $\rho = 0.85 \sigma^{-3}$ for energies in the range
$E = -1.5 \epsilon$ to $E = 2.2 \epsilon$, and for $E = -2 \epsilon$
we studied the cases $\rho = 1 \sigma^{-3}$ and $\rho = 1.09 \sigma^{-3}$.
The temperature
of the periodic system used for constructing the initial configurations
is in the range $\sim 1.4$ to $\sim 4.3\epsilon$. From the equation of
state of the Lennard-Jones system \cite{lj_eos}, it can be seen that
our initial drops are in a hot and compressed.
For each energy we performed and analyzed no less than $120$ explosions,
for $E=1.8 \epsilon$, $E=0.9 \epsilon$, $E=0.5 \epsilon$ and $E= -0.5 
\epsilon$ we studied $300$ evolutions. 

\section{Fragment formation and emission}

As stated in the introduction, many systems, differing greatly
in size, interaction potential, etc, exhibit fragmentation. 
Thus, a good way of characterizing the excitation of the system 
is according to the asymptotic mass spectra. If the energy of the 
system is high, it will break in several small fragments. The 
asymptotic mass spectrum will show rapid decay for large masses. 
On the other hand, for low excitation energies, the system will 
evaporate monomers and small clusters while a big drop, comprising 
most of the mass of the system, will remain bound. In this case the 
mass spectrum is U shaped. A third case is usually found: for a given 
intermediate energy the mass spectrum will show power law behavior. 
This last case is quite important. Taking into account that
a power law implies scale invariance, and power law mass spectra 
are found in second order phase transitions, (e.g. percolation 
at the critical probability \cite{perco} or liquid-gas phase 
transition at the critical point \cite{fisher}), the power law
mass spectra was associated with the system undergoing a second 
order phase transition \cite{bona}. This conjecture, however attractive,
has not been confirmed.  

In Fig. 1 we show the asymptotic mass spectra for six different 
energies, namely $E = 2.4 \epsilon$, $E = 1.8 \epsilon$, $E = 0.9 \epsilon$,
$E = 0.5 \epsilon$, $E =-0.5 \epsilon$, and $E =-2.4 \epsilon$. 
The general behavior described in the last paragraph can be seen
in Fig. 1. Although we are not interested in checking the occurrence of
a second order phase transition, we can see that the power law mass
spectra must be close to $E = 0.5 \epsilon$.

In order to study the mechanisms that lead to fragmentation it is
important to know the time at which the asymptotic fragments form
and the one at which they are emitted.

In previous papers we have fully analyzed the main fragment
recognition algorithms currently in use, see for example \cite{fragments}. 
The simplest definition of cluster
is basically: a group of particles that are close to each other and far away
from the rest. The fragment recognition method known as minimum spanning
tree (MST) is based on the last idea \cite{fragments}. In this
approach a cluster is defined in the following way: given a set of 
particles $i,j,k,...$, they belong to a cluster $C$ if :

\begin{equation}
\forall \hspace{0.2cm} i\in C\hspace{0.2cm} ,\exists \hspace{0.2cm} j\in C
\hspace{0.2cm} /\hspace{0.2cm} \left| {\bf r}_i-{\bf r}_j\right| 
\leq R_{cl}
\end{equation}
where ${\bf r}_i$ and ${\bf r}_j$ denote the positions of the particles and
$R_{cl}$ is a parameter usually referred to as clusterization radius, and
is usually related to the range of the interaction potential. In our
calculations we took $R_{cl} = 3 \sigma$.

On the other hand, the early cluster formation model (ECFM)
\cite{dorso93}, is based on the next 
definition: clusters are those that define the most bound partition of 
the system, i.e. the partition (defined by the set of clusters $\{C_i\}$) 
that minimizes the sum of the energies of each fragment:

\begin{equation}
E_{\left\{ C_i\right\} }=\sum_i\left[ \sum_{j\in C_i}K_j^{cm}+\sum_{j,k\in
C_i}V_{j,k}\right]
\end{equation}
where the first sum is over the clusters of the partition, and $K_j^{cm}$ is
the kinetic energy of particle $j$ measured in the center of mass frame of
the cluster which contains particle $j$. The algorithm (early cluster
recognition algorithm, ECRA) devised to
achieve this goal is based on an optimization procedure in the spirit
of simulated annealing \cite{dorso93}.

It has long been known that the ECRA algorithm finds that the asymptotic 
clusters are formed, in phase space, long before the separate in coordinate 
space, and 
become recognizable with the MST algorithm, i.e. long before they are 
emitted \cite{pbc,times,fragments}. We then associate the time
at which the ECRA method finds the asymptotic clusters to the 
fragment formation time-scale and the one related to the MST analysis 
to the fragment emission time-scale. We will devote the rest of this
section to the calculation of the above mentioned time-scales.

In Fig. 2 we show the time evolution of the mean mass of the greatest
fragments, Fig. 2(a), and the intermediate mass fragments (fragments with
mass in the range $4$ - $50$) multiplicities, Fig. 2 (b), obtained
using both fragment recognition algorithms. It is clear that the
ECRA recognizes the asymptotic fragments before they separate in
coordinate space.

Another important quantity is the microscopic stability of the
clusters.
In order to study this we define the following microscopic 
persistence coefficient. At a given time $t$ the system will be
formed by a set of clusters ${C_i(t)}$ which will become, for
long times, the asymptotic fragments which we will denote ${C_i}$.
Lets consider a given cluster $C_i(t)$ with mass number $n_i(t)$, let
$b_i(t) = n_i(t)(n_i(t) - 1) / 2$ 
be the number of pairs of particles in the cluster. Its 
constituents particles might be at the asymptotic time in two or more
different clusters. We define $a_i(t)$ as the number of pairs of
particles that belong to $C_i(t)$ and also are together in a given asymptotic
cluster. We are now able to define the microscopic persistence coefficient
\begin{equation}
P(t) = \frac{1}{N_{ev}} \sum_{ev} \left[  \frac{1}{\sum_{cl} m_i(t)}
		\sum_{cl} m_i(t) \frac{a_i{t}}{b_i{t}} \right],
\end{equation}
where the first sum runs over the different events for a given
energy, $N_{ev}$ is the number of events, and the other two sums 
run over the clusters at time $t$. It is clear that the persistence coefficient
is equal to $1$ if the microscopic structure of the clusters is equal the
the asymptotic one. On the other hand this coefficient approaches $0$ when
the two partitions under study bear little similarity.
This coefficient can be defined
for ECRA clusters as well as for MST ones. In Fig. 3 we show
$P(t)$ for the ECRA analysis (full lines) and $P(t)$ for the MST 
analysis (dashed lines). In each case the asymptotic partition
was taken as that resulting from the corresponding analysis
(ECRA or MST) for time $t = 150 t_0$. At this ``asymptotic'' time
ECRA and MST analysis yield almost the same results.
The horizontal lines in Fig. 3 represent a reference value related
to an evaporation-like process. It is the value of the persistence coefficient
when each asymptotic partition is compared with itself after removing
one particle from each of it constituents clusters of mass number
greater than $2$, i.e.:
\begin{equation}
P_{ref} = \frac{1}{N_{ev}} \sum_{ev} \left[  \frac{1}{\sum_{cl} 
		m_i(t_{\infty})} 
		\sum_{cl} m_i(t_{\infty}) 
		\frac{m_i(t_{\infty})(m_i(t_{\infty})-1)}
		{(m_i(t_{\infty})-1) (m_i(t_{\infty})-2) } \right],
\end{equation}
where the second sum runs over clusters of mass $3$ and above and 
$m_i(t_{\infty})$ denotes the mass number of cluster $i$ which belongs
to the asymptotic partition.

We can define, from the results shown in Figs. 2 and 3, two different
time scales: a ``break-up or fragment formation time'' $\tau_{ff}(E)$ 
related to the ECRA partition differing from the asymptotic partition
by an evaporation-like process, and a ``fragment emission time'' 
$\tau_{fe}$ given by the time at which the asymptotic fragments separate 
from each other, and consequently the MST partition differs from
the asymptotic one by an evaporation-like process.
We obtain the following time-scales: $\tau_{ff}(1.8\epsilon) \sim 
20 t_0$, $\tau_{ff}(0.9\epsilon) \sim 35 t_0$, 
$\tau_{ff}(0.5\epsilon) \sim 52 t_0$, 
$\tau_{ff}(-0.5\epsilon) \sim 75 t_0$.
It is important to note that at the above times the system is 
already broken in phase space
(ECRA analysis) but not fully broken in coordinate space (MST
analysis), and the biggest MST cluster contains more than half the
total mass of the system,
see Figs. 2 and 3. This last fact will turn out to be important
in order to characterize the break-up state of the system. 

We obtained the following time-scales for fragment emission:
$\tau_{fe}(1.8\epsilon) \sim 
40 t_0$, $\tau_{fe}(0.9\epsilon) \sim 60 t_0$, 
$\tau_{fe}(0.5\epsilon) \sim 67 t_0$, 
$\tau_{fe}(-0.5\epsilon) \sim 100 t_0$.


\section{Internal temperature of the fragments}

The internal state of the asymptotic clusters is a very important
quantity in fragmentation. The reason is twofold, on the one
hand the asymptotic excitation of the clusters is accessible
experimentally in nuclear fragmentation \cite{serfling} and in 
small atomistic clusters \cite{calorimetria}. On the other hand, 
as we will show, the internal temperature of the asymptotic clusters
gives information about the break-up state of the system. We
calculate the internal temperature of the clusters of mass number
$n>1$ in the following way:
\begin{equation}
T_{cl}(n) = \frac{1}{N(n)} \sum_{j} 
	\frac{1}{3n - 3} \sum_{i \in j} \frac{1}{2}m (v_i^{j-cm})^2,
\end{equation}
where $N(n)$ is the number of clusters of mass number $n$ in all events
for a given energy, the first sum runs over all clusters $j$ of
mass $n$, the second sum runs over the particles $i$ belonging to
cluster $j$ and $v_i^{j-cm}$ denotes the velocity of particle $i$
measured from the c.m. frame of the cluster it belongs to.

In Fig. 4(a) we show the internal temperature of the asymptotic clusters 
($T_{cl}(n)$) as a function of its mass number for the wide range of 
energies considered; in
Fig. 4(b) the same quantity is plotted but at fragment formation time
($\tau_{ff}(E)$), in this last case we used the ECRA clusters. Fig. 4 
shows a very important result: the temperature
of the asymptotic clusters depends only on their mass, and not on the
total energy of the fragmenting system, furthermore at break-up time 
$\tau_{ff}$
the temperature of the fragments is basically equal to that of the
asymptotic ones. This means that at break-up time the clusters have 
already cooled down to a long lived metastable state in which they need no 
further relaxation, except for little evaporation. 
For times greater than $\tau_{ff}(E)$, the clusters simply fly away from each
other but nothing really important happens regarding their internal degrees 
of freedom, except, again, for some evaporation. The same behavior 
was found for
two dimensional Lennard Jones drops \cite{caloric}. 
Let us mention that the energies shown in Fig. 4, represent very different
behaviors regarding the fragmentation pattern, it can be seen that
for $E = 1.8\epsilon$ there are no clusters of mass bigger than
$\sim 30$, while for $E = -0.5 \epsilon$ there are little intermediate
mass fragments, which comes from a U-shaped mass spectra.

\section{Energy partition: collective expansion, internal kinetic and
potential energies}

In order to study the development of the collective flux and the
way the total energy is partitioned we take the following approach.
Since the radial collective velocity is position dependent,
the outer particles expand at a faster rate than the inner ones,
we divide our drops in concentric spherical regions.
The $i^{th}$ region, is formed by the points ${\bf r}$ in coordinate 
space that satisfy:

\begin{equation}
	\delta r \, (i-1) \le \left| {\bf r} \right| < \delta r\, i
\end{equation}
where $\delta r $ is the width of the regions, we took $\delta r = 2
\sigma$. ${\bf r}$ is measured form the c.m. of the system, as will be 
any coordinate position throughout the paper. 

We now define the mean radial velocity of region $i$ as:
\begin{equation}
	v_{rad}^{(i)}(t) = \frac{1}{N_i(t)} \sum_{ev} \sum_{j \in i} 
		\frac{ {\bf v}_{j}(t) \cdot {\bf r}_{j}(t)}
			{\left| {\bf r}_{j}(t) \right|}
\end{equation}
where the first sum runs over the different events for a given energy,
the second over the particles $j$ that belong, at time $t$, to region
$i$ and ${\bf v}_j$ and ${\bf r}_j$ are the velocity and position of
particle $j$. $N_i(t)$ is total number of particles belonging to
region $i$ in all the events.

We can divide the total kinetic energy per particle in two parts: 
\begin{equation}
	K(t) = K_{coll}(t) + K_{int}(t).
\end{equation}
The term $K_{coll}(t)$ is related to the collective motion:
\begin{equation}
	K_{coll}(t) = \frac{1}{n_{ev} N} 
			\sum_{regions} N_i(t) \frac{m}{2} 
			\left( v_{rad}^{(i)}(t) \right)^2
\end{equation} 	
where $m$ is the mass of each particle, $n_{ev}$ is the number of
events and $N=147$ is the number of particles in each drop. 
The second term, $K_{int}(t)$ is related to the ``internal'' kinetic 
energy:
\begin{equation}
	K_{int}(t) = \frac{1}{n_{ev} N}
		\sum_{regions} \sum_{j \in i} 
			\frac{1}{2} m 
			\left( 
	{\bf v}_j - v_{rad}^{(i)} \cdot {\bf \hat{r}}_j
			\right)^2
\end{equation}
where ${\bf \hat{r}}_j = {\bf r}_j / |{\bf r}_j|$.

In Fig. 5 we show $K(t)$, $K_{coll}(t)$, $K_{int}(t)$ and
the potential energy, $V(t)$, as a function of time, for
four different cases. During the initial stage of the fragmentation
process the collective expansion builds up, while the internal
kinetic energy diminishes, i.e. the system relaxes. The potential
energy grows in time because the system is fragmenting and consequently
increasing its surface.

In order to see the way in which the mean radial velocity
depends on position, we show in Fig. 6 the radial velocity profiles
$v_{rad}^{(i)}(\tau_{ff})$ at break-up time, for the same energies as in 
Fig. 5. The position dependence is clear and so is the fact that the
expansion velocity increases with the total energy.

What we defined as the internal kinetic energy can be related
to ``local temperature'' which is usually defined as the velocity
fluctuations around the mean collective velocity, which in our case
is the mean radial velocity \cite{huang}, i. e.:
\begin{equation}
T_{loc}^{(i)} = \frac{1}{N_i(t)} \frac 2 3
         \sum_{j \in i}
            \frac{1}{2} m
            \left(
    {\bf v}_j - v_{rad}^{(i)} \cdot {\bf \hat{r}}_j
            \right)^2,
\end{equation} 

In this way we are making the conjecture
that the fragmenting system is in local equilibrium and
the velocity distribution follows \cite{huang}:
\begin{equation}
    f({\bf v};{\bf r}) = \rho({\bf r})
        \left( \frac{m \beta({\bf r})}{2 \pi} \right)^{3/2} \,
         e^{ \beta({\bf r}) \frac{m}{2}
        ({\bf v} - {\bf v}_{rad}({\bf r}))^2 }
\end{equation}
where  $\rho({\bf r})$ and  $\beta({\bf r})$ are the local density and inverse
of the local temperature respectively; ${\bf v}_{rad}({\bf r})$ 
is the collective
velocity which in our case is in the radial direction.
In Fig. 7 we show the local temperature profiles for different
energies, namely $E = 1.8\epsilon$ (full lines), $E = 0.9\epsilon$
(dotted lines), $E = 0.5\epsilon$ (dashed lines) and $E =-0.5\epsilon$
(dashed-dotted lines), and for different
times. In Fig. 7(a) full symbols denote the local temperature profiles 
at break-up time and empty symbols correspond to the initial
configuration, i.e. $t = 0$ and in Fig. 7(b) full symbols denote, again, 
break-up time and empty symbols correspond to asymptotic times.
In order to check validity of the local 
equilibrium conjecture and the physical meaning of the local temperature 
we analyzed the degree of isotropy of the velocity fluctuations
around the expansion.
We found that the local temperature related to velocity fluctuations
in the radial direction is equal to that related to transverse
fluctuations. As a consequence the expanding system can be considered to
be in local equilibrium at $\tau_{ff}$. 

From Fig. 7 we can clearly see
that while the initial temperature profiles are quite different
in the cases shown (corresponding to very different excitation energies)
the local temperature profiles at break-up time are quite similar in 
all the cases. In this way we can characterize the velocity distribution
at the break-up state as a mean radial velocity, which grows almost
linearly with distance from the c.m. of the system, and
which depends on the total energy of the system plus velocity fluctuations,
on top of the collective motion, which do not depend 
on the total excitation of the system. This result is important because
it represent the first step in the characterization of the break-up
state. At this point the central problem is understanding why the local 
temperature at break-up is the same for all initial conditions (i.e.
excitation) and what is the meaning of its value. We will devote next
section to answer that question. It is also interesting to note
the no evidence of ``gas phase'' phase behavior, i.e. the increase 
of temperature for high energies, is found for the local temperature.

In Fig. 8 we show the time evolution of the local temperature averaged
over the three innermost regions (which comprise approximately
the volume of the initial drop) for different energies.
It can be seen that during the first stage of the fragmentation 
process part of the initial thermal energy of the system is converted 
to collective energy and the local temperature decreases. This process
continues until a given temperature is achieved, this value of
temperature is quite independent of the total energy of the system,
but it can be seen that it slowly diminishes as the total energy
increases. It is worth mentioning at this point that the energies
shown in Fig. 8 represent very different behaviors of the system
regarding its asymptotic states, see the mass spectra in Fig. 1.
A similar behavior was found for classical two dimensional drops
\cite{caloric}.

We now focus our attention on the way the total energy of the system is
partitioned.
The partition of energy as a function of the total energy of the
system is undoubtedly important for the understanding of the
process of fragmentation. in Fig. 9(a) we show the asymptotic values
of collective and internal kinetic energy, the potential energy 
and the average local temperature of the three inner most region as a 
function of the total energy.
Perhaps more important that the asymptotic values are those at
the break up time. In Fig. 9(b) we show the same quantities, see caption
for details, but for the break-up time $\tau_{ff}(E)$.
It can clearly be seen that the local temperature at break-up
is quite independent of the total energy, in contrast with the
other quantities that vary noticeably in the energy range shown.
Furthermore it is very interesting to note that this value
of the local temperature is very similar to the internal temperature
of the greater clusters, see Fig. 4.
The meaning of the constancy of the local temperature and of its value
will become apparent in the next section.

As already mentioned in the introduction we can define a cluster 
local temperature related to the fluctuations of the c.m. velocity 
of the clusters over the collective expansion:
\begin{equation}
	T_{cl loc}^{(i)}(t) = \frac{2}{3 N_i(t)}
			\sum_{j \in i}
			\frac{1}{2} m 
			\left( 
	{\bf v}_{cm}^{(j)} - \frac{v_{rad}^{(i)} \cdot {\bf r}_{c.m.}^{(j)}}
		{\left| {\bf r}_{c.m.}^{(j)}  \right|}
			\right)^2
\end{equation}
where the sum runs over the clusters whose c.m. position belongs to the
$i^{th}$ region in all the events, ${\bf v}_{cm}^{(j)}$ is the c.m. 
velocity of cluster
$j$ and ${\bf r}_{cm}^{(j)}$ is, of course, the position of the c.m.
of the cluster. We will use the name ``local temperature'' for the
one related to fluctuations of the particle velocities and
``cluster local temperature'' for the one defined in the last
equation.

In Fig. 10 we show the cluster local temperature profiles for energies
$E = 1.8\epsilon$, $E = 0.9\epsilon$, $E = 0.5\epsilon$ and 
$E =-0.5\epsilon$ at break-up time. The profiles are similar to
those of the local temperature, Fig. 7. Taking into account that
the number of clusters is much smaller than the number of particles
the values plotted are more likely to
deviate from their mean value. This quantity is very important because
contrary to the local temperature it can be measured in experiments,
at least at asymptotic times \cite{pocho97}. Again, no evidence of the
``gas branch'' is found.

\section{Extended caloric curve}

In this section we show the caloric curve for our 3 dimensional 
L-J drops, Fig. 11, in a very wide energy range. The ``extended
caloric curve'', as we will name it, encompasses the solid-like phase
(region I), the liquid-like phase (region III), the associated phase 
transition (region II), and also the higher energy process of 
evaporation and multifragmentation (region IV).

The low energy part of the caloric curves (regions I, II and III)
was obtained by analyzing MD simulations whose initial configurations
were obtained by rescaling the velocities of a $N=147$ particles drop
originally constructed close to its ground state. This low energy 
configuration was obtained simply by cutting a spherical drop from a FCC 
crystal, whose density was taken as the close-packed one ($\rho \sim 1.09
1/\sigma^3$) and whose temperature was close to zero. 
Once this initial configuration was cut the particle velocities were set
to zero and the system was evolved for a long time to achieve
thermalization. In order to study the drop for different energies,
the velocities of the thermalized low energy system were rescaled
so as to get the desired total energy. At the low energies in
the range encompassed by regions I, II and III of Fig. 11 the
LJ system is self confined except for some little evaporation
at the higher energies of region III.
The temperature in regions I, II and III is related to c.m. kinetic 
energy of the drop, disregarding any evaporated particle. 

In region IV we plot the local temperature averaged over the three 
inner most regions from our fragmentation computer experiments,
also averaged over a time $t = 20t_0$ centered at the break-up time,
see Fig. 7.
This means that the extended caloric curve in the fragmentation
region represents the break-up temperature of the system.
Let us mention that at break-up time the three central regions
contain $\sim 60$ to $\sim 90$ particles, and a big, interacting, 
MST cluster is still present, see Fig. 2.

Regions I, II and III feature the well known solid-like to 
liquid-like phase transition in small systems \cite{labastie90}. 
In region I the drop is solid, the solid-liquid phase transition 
appears as the loop in region II. In region III the drop is liquid-like. 
In this low energy regime the behavior of the drops resemble that of the
macroscopic systems, although there are some important differences,
see \cite{labastie90}.
It is clear that an isolated liquid drop cannot be heated without limits.
Once a certain temperature is attained, which depends on the size
of the system, if more energy is supplied to the system it
will evaporate particles but it will not heat up; we will call
this temperature the limit temperature $T_{lim}$. This feature
can be seen in the high energy region of region III. For energies
higher than that of the evaporating liquid the system undergoes the non
equilibrium process of fragmentation, region IV of
Fig. 11. Of course the limit between evaporation and fragmentation
(regions III and IV) is arbitrary. Due to this ``overlap'',
for $E = -2 \epsilon$ we performed evolutions applying the velocity
rescaling method used for studying low energy drops and also constructing
the initial configurations from the periodic system, i.e. the method
used to study fragmentation. It is important to notice from Fig. 11, 
$E= -2\epsilon$ that we obtained the same value of temperature 
regardless of the way we constructed the drop and the definition of 
temperature used.

It is worth mentioning again that the caloric curve in the fragmentation
region represent the local temperature of the central regions of
the expanding system at break-up. This process appears as a quite constant
temperature region of the caloric curve, decreasing slowly for
high energies.
It can be seen from Fig. 11 and from the temperature profiles, Fig. 7, 
that the local temperature at break-up is quite independent of the 
total energy of the fragmenting system. Note that the initial drops in 
our fragmentation computer experiments (described in section II)
are artificially constructed hotter than the limit 
temperature, in real cases this state is achieved via a sudden input 
of energy like in collisions. In its evolution the system cools
down, while the expansion builds up, until it reaches the limit
temperature, which of course does not depend on the initial excitation
of the system, and needs no further relaxation. From this time on
the system continues its expansion at constant velocity and
the temperature remains quite constant. The slow decrease of the
temperature for high energies is related to the fact that the
size of the drops diminishes (as the system breaks into more
fragments) and consequently their limit temperature decreases too.
Within this picture we can understand why the internal temperature
of the fragments, from $\tau_{ff}$ on, is independent of the initial
excitation of the system, Fig. 4. The system expands until it needs
no further relaxation, which means that the fragments will be ``as hot
as they can''. Notice that the internal temperature of the big clusters
is very similar to the limit temperature of the $N=147$ drop. 

\section{Discussion}

The fragmentation process can be divided in three stages. The first
one which we will call flux and fragment formation stage, goes from
$t = 0$ to $t = \tau_{ff}(E)$. During this stage the radial flux
and density fluctuations develop, these elements determine the cluster
partition according to the ECFM model. By the end of this stage
the asymptotic fragments are already formed although most of the mass 
of the system is still interacting and forming a big configurational
(MST) cluster. During this stage and while the collective motion
develops the system cools down. We showed that some degree of local
equilibrium is achieved and that a local temperature can be defined. 
At $\tau_{ff}$ the local temperature profiles are quite independent 
of the total energy of the system, of course the initial profiles
depend on the total energy of the system and on the initial density.
The value of the local temperature in the central region of the drop
is equal to the limit temperature of its constituents clusters, i.e.
the maximum internal temperature that the liquid clusters can have.
This means that the initial cooling process continues until the 
temperature attains
its limit value and the system needs no further relaxation. 
This is why the local temperature at break-up is quite independent
of the total energy, and it
slowly decreases when the total energy increases; as the total energy of
the system increases it breaks into smaller clusters and consequently the
limit temperature diminishes.

The second stage of the fragmentation process, which we will name
fragment emission stage, goes from $\tau_{ff}(E)$ to $\tau_{fe}(E)$. 
During this stage the already formed fragments are emitted,
i.e. they separate in configurational space and become recognizable with
a MST cluster analysis. The local temperature profiles are quite constant
during this second stage (Fig. 7(b)) and so is the internal temperature 
of the clusters, Fig. 4. Most of the cooling and relaxation has already 
taken place during the fragment and flux formation stage.

The third stage encompasses times greater than $\tau_{fe}$ and will be
called free expansion stage. During this stage the already emitted 
clusters expand freely. Only some evaporation may occur.

We have also studied the cluster local temperature, related to
the clusters c.m. velocity fluctuation over the expansion. This 
quantity is important because it can be measured in experiments
if the collective expansion is precisely identified and subtracted.
The cluster local temperature profiles are similar to the local
temperature ones. In this way we see that different degrees of freedom
are quite thermalized and give similar temperatures. 
Let us recall once more that in nuclear fragmentation experiments
some results show a caloric curve featuring a ``gas branch'' while 
others do not;
this discrepancy might be explained, as proposed in \cite{pocho97},
by considering that different degrees of freedom freeze out at
different temperatures.
For our classical system the internal degrees of freedom of
the clusters and their c.m. velocity fluctuations around the
mean radial velocity yield very similar values of temperatures.  
The origin of the discrepancies in the experimental nuclear caloric
curve might be another; like not taking in account the collective
expansion properly in the temperature definition.

The results found for the different temperatures are consistent
with our local equilibrium conjecture. If the particle velocity
distribution followed:
\begin{equation}
    f({\bf v};{\bf r}) = \rho({\bf r})
        \left( \frac{m \beta({\bf r})}{2 \pi} \right)^{3/2} \,
         e^{ \beta({\bf r}) \frac{m}{2}
        ({\bf v} - {\bf v}_{rad}({\bf r}))^2 }
\end{equation}
all the temperatures defined in this work should yield the same
value which is precisely what we found. This is very important
fact, because it means that we can know the break-up temperature
by measuring the internal temperature of the asymptotic clusters
or the cluster local temperature, if the collective motion is
properly taken care of, at least for systems similar to the one
considered in this study.

Taking the above mentioned into account, we have calculated the extended 
caloric curve
that describes the thermal behavior of our L-J drop from the solid-like
regime all the way up to the fragmentation regime. The resulting picture
shows the standard behavior up to the liquid-like state. When the energy
is further increased leading to evaporation and fragmentation the 
break-up temperature
displays a plateau followed by a monotonous, slow, decrease as the system
fragments into smaller and smaller clusters, similar behavior
was found in \cite{serfling,hauger}. In macroscopic systems a
plateau in the caloric curve denotes a phase transition, for example
in the solid-liquid phase transition an input of energy will not result
in an increase of temperature but in melting. A similar process happens
in the solid-like to liquid-like phase transition in small drops 
\cite{labastie90}.
In the case of fragmentation, if the energy of the initial condition
is increased the collective expansion will grow, the system will
break into smaller fragments but the break-up temperature will
not increase.

It is worth mentioning at this point that none of the temperature
definitions that we studied show any kind of evidence of the
``gas branch'' from the fragmentation computer experiments, which 
appears as a steady increase of the temperature in 
the caloric curve, found for high energies in \cite{pocho95}. 
In our calculations, the cluster internal 
temperature, the cluster local temperatures and the local temperature slowly 
decrease as the total energy of the system increases. 
This can be understood, within our 
multifragmentation picture taking into account the relation 
between the temperature of the fragmenting system with the limit
temperature of the asymptotic clusters. The remaining energy is converted
into collective kinetic energy. The presence of this collective
motion is responsible for the break-down of the standard picture
of liquid-gas phase transitions. Within our picture of fragmentation
no ``gas phase'' behavior is to be expected, i.e. the temperature will
not increase with energy but decrease as the system expands in a more
orderly way.

\acknowledgements
This work was done under partial financial support from the
University of Buenos Aires via grant EX-070.

\figure{FIG. 1. Asymptotic mass spectra for different initial conditions.
(a) $E = 2.2\epsilon$, (b) $E = 1.8\epsilon$,
(c) $E = 0.9\epsilon$, (d) $E = 0.5\epsilon$,
(e) $E =-0.5\epsilon$, (f) $E =-2\epsilon$. In the cases (a), (b), (c)
(d) and (e) the initial density is $\rho = 0.85 \sigma^{-3}$ and in 
(f) it was $\rho = 1 \sigma^{-3}$.
 }

\figure{FIG. 2. (a) Intermediate mass fragments multiplicities as a
function of time for different energies. (b) Mean maximum cluster
mass vs. time. Dashed lines join points obtained with the ECRA method
while full lines join MST points.
Circles denote $E = 1.8\epsilon$, squares denote $E = 0.9\epsilon$,
diamonds denote $E = 0.5\epsilon$ and triangles denote $E =-0.5\epsilon$.
}

\figure{FIG. 3. Microscopic persistence coefficient as a function of
time for different cases. Full lines denote ECRA results and dashed 
lines denote MST
results. The horizontal lines indicate the evaporation reference,
see text. 
(a) $E = 1.8\epsilon$, (b) $E = 0.9\epsilon$,
(c) $E = 0.5\epsilon$ and (d) $E =-0.5\epsilon$.
}

\figure{FIG. 4. Internal temperature of the clusters as a function
of their mass number, for asymptotic times (a) and at break-up time
(b). 
Full circles denote $E = 1.8\epsilon$, squares denote $E = 0.9\epsilon$,
diamonds denote $E = 0.5\epsilon$ and triangles denote $E =-0.5\epsilon$.
}

\figure{FIG. 5. Total (full lines), collective (dashed-dotted lines)
and internal (dashed lines) kinetic energy, and potential energy
(dotted lines) as a function of time for different cases. 
(a) $E = 1.8\epsilon$, (b) $E = 0.9\epsilon$,
(c) $E = 0.5\epsilon$ and (d) $E =-0.5\epsilon$.
}

\figure{FIG. 6. Radial velocity profiles at break-up time.  
Circles denote $E = 1.8\epsilon$, squares denote $E = 0.9\epsilon$,
diamonds denote $E = 0.5\epsilon$ and triangles denote $E =-0.5\epsilon$.
}

\figure{FIG. 7. (a) Initial (empty symbols) and break-up (full symbols)
local temperature profiles. 
(b) Break-up (full symbols) and asymptotic (empty symbols)
local temperature profiles. 
Full lines denote $E = 1.8\epsilon$, dotted lines denote $E = 0.9\epsilon$,
dashed lines denote $E = 0.5\epsilon$ and dashed-dotted lines
 denote $E =-0.5\epsilon$.
}

\figure{FIG. 8. Average local temperature of the three innermost
regions as a function of time.
Dashed-dotted lines denote $E = 1.8\epsilon$, dotted lines denote 
$E = 0.9\epsilon$,
dashed lines denote $E = 0.5\epsilon$ and full lines
 denote $E =-0.5\epsilon$.
}

\figure{FIG. 9. Energy partition. Total (full lines), collective
(dashed-dotted lines) and internal (dashed lines) kinetic energy, 
potential energy (dotted lines) and average local temperature of
the three innermost regions (full thick lines) as a function of
the total energy of the system. (a) At asymptotic times and (b) at
break-up time.
}

\figure{FIG. 10. Cluster local temperature profiles. Cluster local
temperature as a function of distance from the c.m. of the system.
Full circles denote $E = 1.8\epsilon$, squares denote $E = 0.9\epsilon$,
diamonds denote $E = 0.5\epsilon$ and triangles denote $E =-0.5\epsilon$.
}

\figure{FIG. 11. Extended caloric curve $T(E)$. Regions I, II and III 
come from the equilibrium simulations. Region IV denotes fragmentation
and the average local temperature of the three innermost regions
at break-up is plotted.}

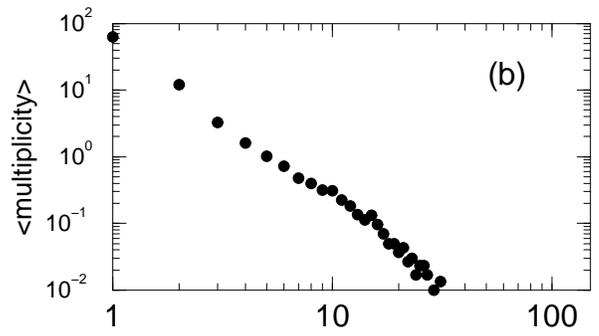
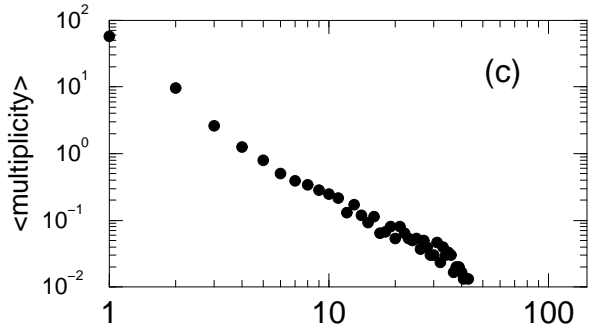
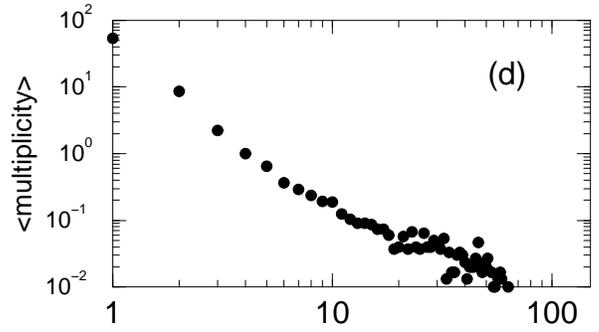
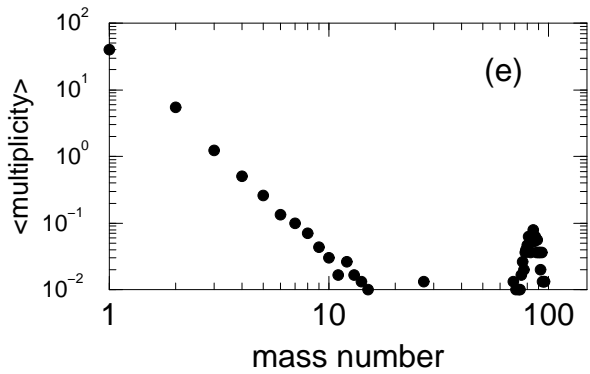
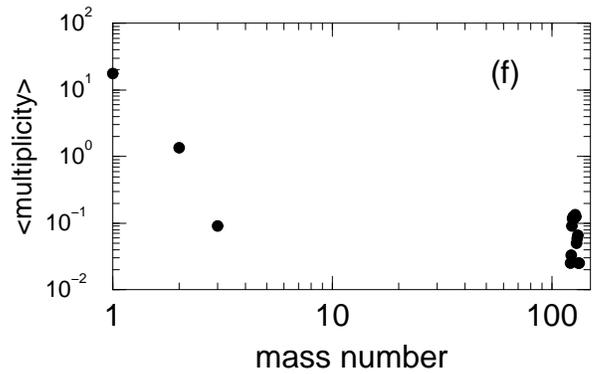
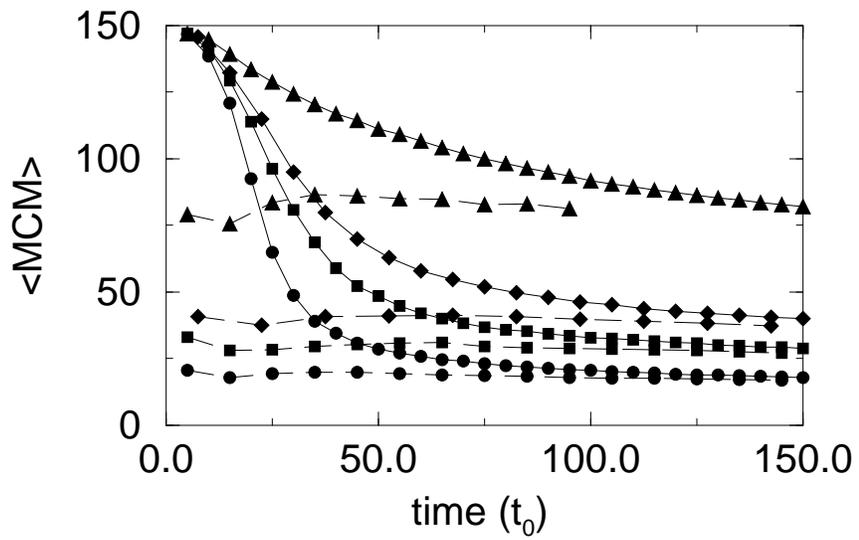
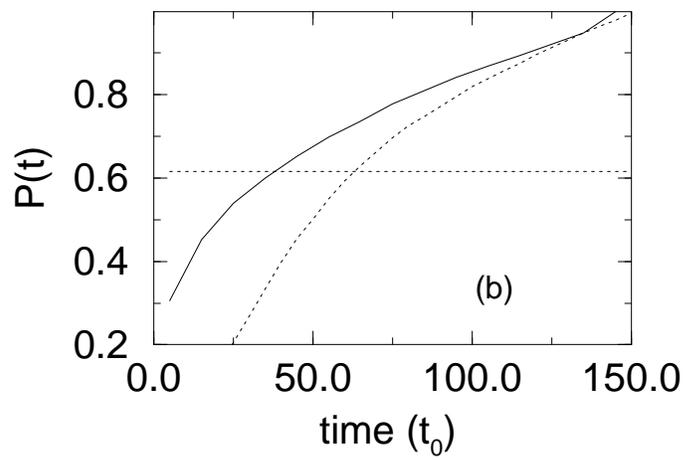
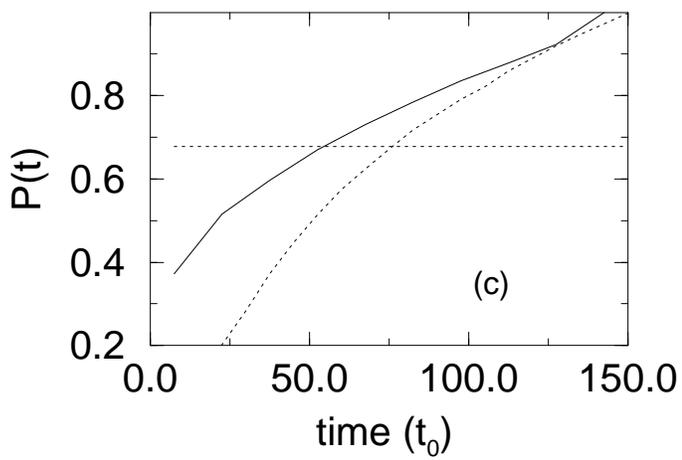
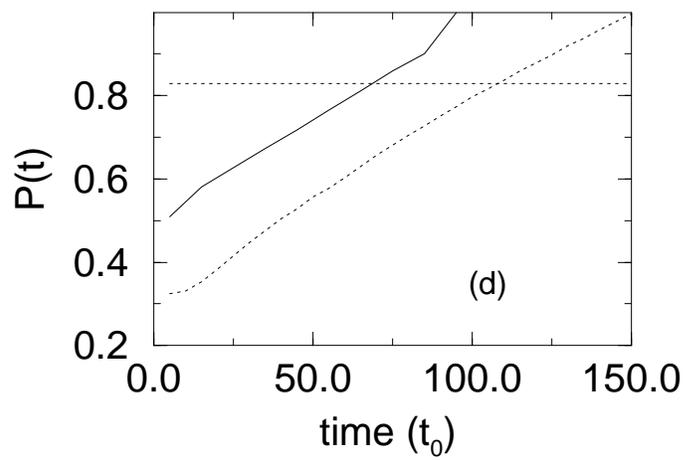
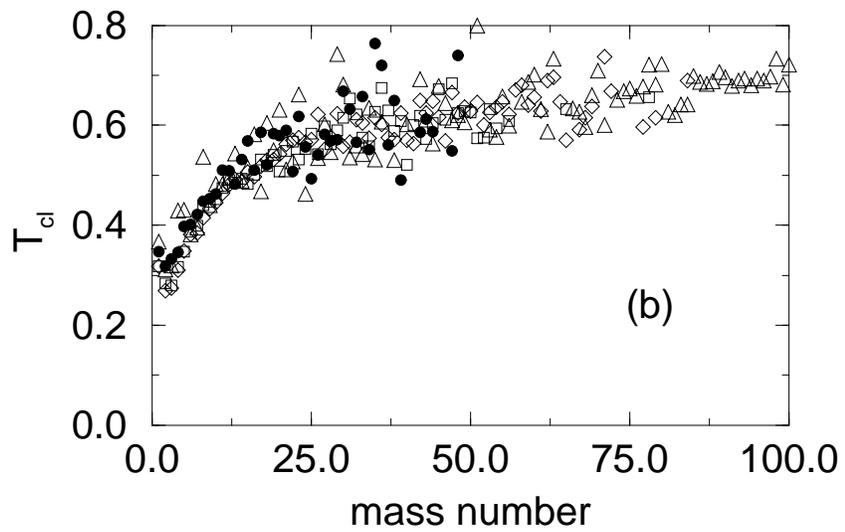
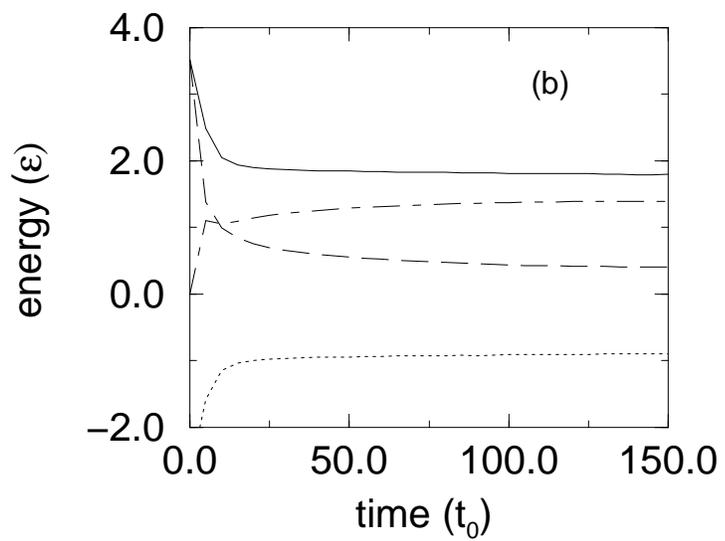
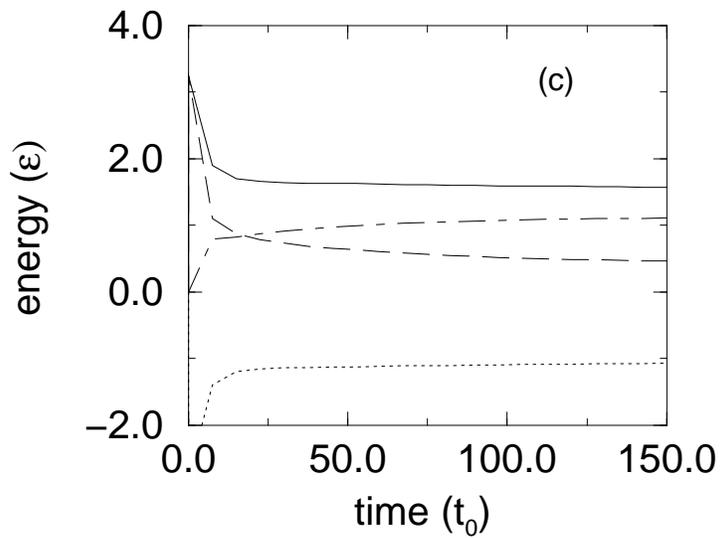
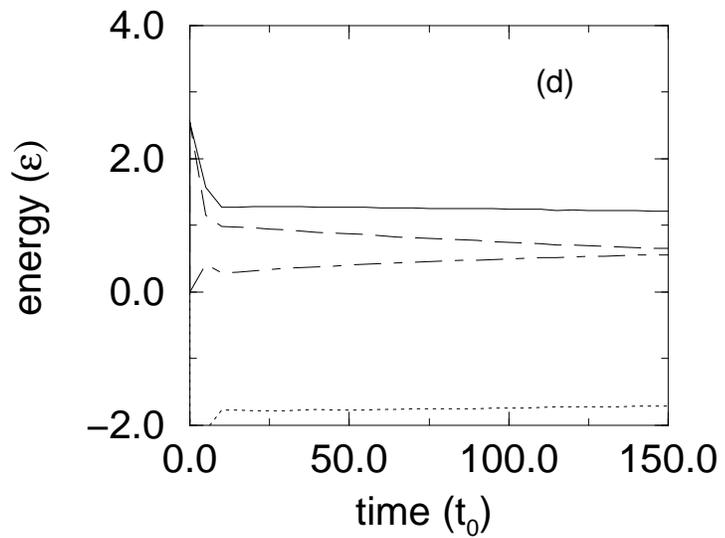
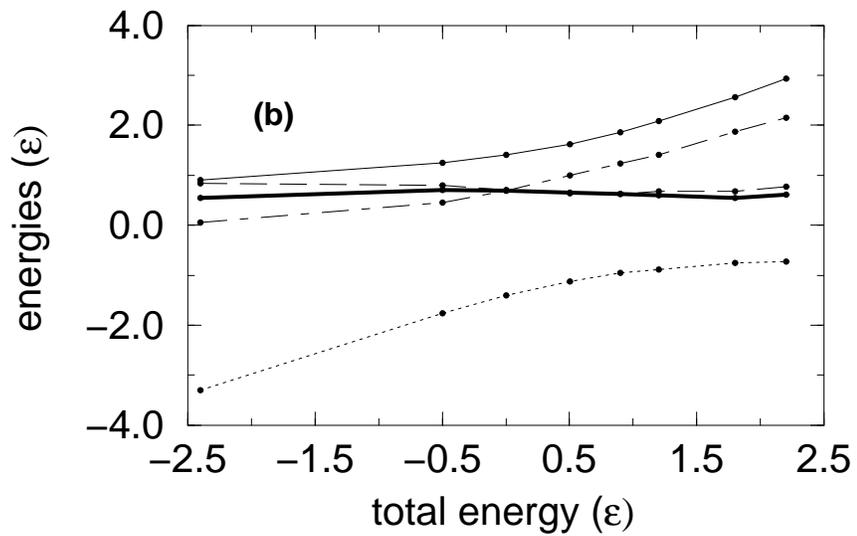
\end{document}